\documentclass[conference]{IEEEtran}

\usepackage{cite}

\usepackage{url}

\usepackage[dvips]{graphicx}

\usepackage[cmex10]{amsmath}

\usepackage{array}

\usepackage[caption=false,font=footnotesize]{subfig}

\begin{document}

\title{Coupled Graphical Models and Their Thresholds}

\author{
  \IEEEauthorblockN{S. Hamed Hassani,  Nicolas Macris and Ruediger Urbanke}
  \IEEEauthorblockA{Laboratory for Communication Theory\\
    School of Computer and Communication Science\\
    Ecole Polytechnique F\'ed\'erale de Lausanne\\
    EPFL-IC-LTHC-Station 14, CH-1015 Lausanne, Switzerland\\
    Email: \{hamed.hassani, nicolas.macris, ruediger.urbanke\}@epfl.ch}
}


\maketitle

\begin{abstract}
\boldmath
The excellent performance of convolutional low-density parity-check
codes is the result of the spatial coupling of individual underlying
codes across a window of growing size, but much smaller 
than the length of the individual codes. Remarkably, the belief-propagation threshold of the coupled
ensemble is boosted to the maximum-a-posteriori 
one of the individual system. We
investigate the generality of this
phenomenon beyond coding theory: we couple general graphical models into
a one-dimensional chain of large individual systems. For the later we take
the Curie-Weiss, random field Curie-Weiss, $K$-satisfiability, 
and $Q$-coloring models.
We always find, based on analytical as well as numerical calculations,
that the message passing thresholds of the coupled systems come very
close to the static ones of the
individual models. 
The remarkable properties of convolutional low-density parity-check codes are a manifestation of this very general phenomenon.
\end{abstract}



\section{Introduction}

Convolutional low-density parity-check (LDPC) codes 
initialy introduced by Felstr\"om and 
Zigangirov \cite{Felstrom-Zizangirov} have been recognized to have 
excellent performance, and have spurred a large body of work (see \cite{Engdahl-Zizangirov},
\cite{Lentmaier-Trubachev-Zizangirov}, 
\cite{Tanner-Sridhara-Sridhrara-Fuja-Costello}, 
\cite{Lentmaier-Fettweis-Zizangirov-Costello} and references in \cite{Kudekar-Richardson-Urbanke-II}). 
A complete mathematical analysis of the mechanism which 
operates behind these constructions has been achieved 
recently 
\cite{Kudekar-Richardson-Urbanke-II} for the binary erasure 
channel (BEC).
Convolutional LDPC ensembles are constructed by coupling together, 
across a window of finite width, copies of a 
standard individual LDPC ensemble 
into a one dimensional chain. 
In order to sucessfully 
decode, one typicaly "terminates" 
the chain by assuming that the 
codebits at the two ends of the chain are known to the decoder.

\begin{figure}
\begin{centering}
\input{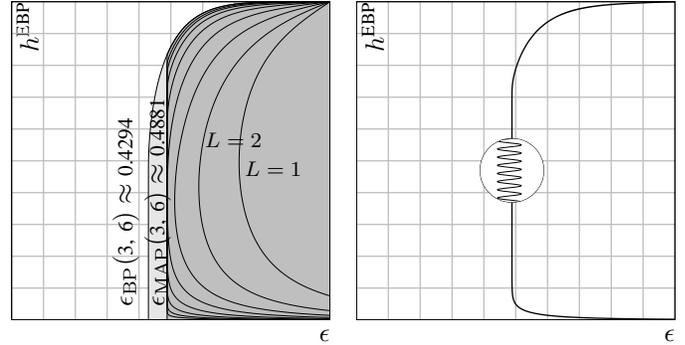}
\caption{{\small Taken from \cite{Kudekar-Richardson-Urbanke-II}.
{\it Left:} EBP EXIT curves of the 
ensemble $(3, 6, L)$,  $L=1, 2, 4, ...,128$ and nearest neighbor coupling.
The light/dark gray areas mark the interior of 
the BP/MAP EXIT function
of the underlying $(3, 6)$-regular ensemble. {\it Right:}
circle shows a magnified portion of the EBP EXIT curve 
for the $(3, 6, 32)$ ensemble. The horizontal magnification is $10^3$.
}}
\label{gexit-bec}
\end{centering}
\end{figure}

The natural thresholds involved are $\epsilon_{\rm BP}$, $\epsilon_{\rm MAP}$ 
for the belief propagation (BP) and maximum-a-posteriori (MAP)
decoders of the individual LDPC ensemble and the ones 
of the coupled ensemble 
$\epsilon_{\rm BP}^{\rm coupled}$, $\epsilon_{\rm MAP}^{\rm coupled}$.
When the length of 
the chain tends 
to infinity, 
$\epsilon_{\rm MAP}^{\rm coupled}\to\epsilon_{\rm MAP}$ from above. 
But the main reason for the
success of convolutional LDPC ensembles is 
that as the width 
of the window 
increases (the size of the individual ensemble 
and the length of the chain being already 
very large) $\epsilon_{\rm BP}^{\rm coupled}\to \epsilon_{\rm MAP}$.
In fact more is true. 
Figure \ref{gexit-bec} shows (BEC channel) how the EBP EXIT curves
of chain ensembles of various lengths,
approach the MAP EXIT curve of the individual 
ensemble \cite{Kudekar-Richardson-Urbanke-II}. 
In the limit, the difference 
between the two curves is indistinghuishable 
even for a small coupling width. These features have also been observed, and partially proved, for general channels, and provide a new way to construct capacity achieving codes.

In this paper we show that the mechanism that is at work is very natural and general. 
We investigate a wide variety of individual graphical models that we couple together to form a one dimensional chain with appropriate boundary conditions. The individual models include the Curie-Weiss (CW) and random field Curie-Weiss (RFCW) models of statistical mechanics, which are defined on
the complete graph, as well as constraint satisfaction problems such as $K-{\rm SAT}$ ($K$ satisfiability) and $Q-{\rm COL}$ ($Q$ coloring) formulated on random sparse graphs. As will be argued later, the underlying graph should be "infinite dimensional" so that the individual system displays a "mean field" behaviour.
We provide analytical arguments as well as numerical evidence that in all these cases, as the width of the coupling increases, the message passing threshold 
of the chain comes close - in fact converges - to the static threshold of the individual system. 

We first discuss the coupled CW models which can be solved, to a large extent, by explicit analytical calculations and thus provide a nice pedagogic illustration of the main mechanisms. We then turn to the more difficult constraint satisfaction problems. 


\section{Coupled Models on Complete Graphs}

The CW model is perhaps the simplest mean field model. It consists of an Ising system of $N$ spins $s_i=\pm 1$ attached to the vertices of a complete graph. The Hamiltonian function is 
$H_N({\underline s}) = -\frac{J}{N}\sum_{\langle i,j\rangle} s_i s_j$
where $J>0$ and the sum carries over all $\frac{N(N-1)}{2}$ 
edges $\langle i,j\rangle$ of the graph. 

\begin{figure}
\begin{center}
\input{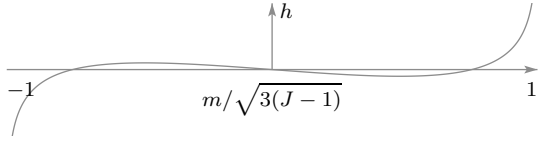}
\end{center}
\caption{{\small Van der Waals curve for the CW model at $J=1.1$. The static phase transition threshold given by the Maxwell construction is $h_s=0$. 
The local maximum and minmum at heights 
$\pm h_{\rm it}$ are iterative thresholds.}}
\label{VdW}
\end{figure}

The easiest way to solve this model is to consider the canonical ensemble in which the free energy is  
$\Phi(m) = -\lim_{N\to +\infty}\frac{1}{N}\ln Z_N(m)$, with the canonical partition function $Z_N(m)$ 
equal to the sum of the Gibbs weights $\exp(- H(\underline s))$ over all spin configurations with 
their magnetization fixed to $mN$, i.e.  $\sum_{i=1}^N s_i =mN$.
A standard calculation gives (up to an irrelevant constant)
$\Phi(m)=-\frac{ J}{2}m^2 - \mathcal{H}(m)$
where the first term is the internal energy of the spin configurations and the second 
one is the contribution from their entropy $\mathcal{H}(m)=-\frac{1+m}{2}\ln \frac{1+m}{2} - \frac{1-m}{2}\ln \frac{1-m}{2}$. 
Here $\Phi(m)$ is not convex because of the non local interaction, and the  thermal equilibrium (Helmoltz) 
free energy is given by the convex hull of $\Phi(m)$. From the thermodynamic relation 
between the magnetic field and the magnetization $h=\frac{\partial\Phi(m)}{\partial m}$ one gets the 
so-called Van der Waals curve (see figure \ref{VdW})
\begin{equation}\label{vdw}
h = -Jm + \frac{1}{2} \ln \frac{1+m}{1-m}\,,
\end{equation}
which is equivalent to the CW equation 
$m=\tanh(Jm +h)$.
The isotherm (i.e. the relation between the  magnetic field and the magnetization at thermal equilibrium) is
 not given by the full Van der Waals curve but by the Maxwell equal area construction which yields $h_s=0$ for the phase transition threshold. The part of the Van der Waals curve not on the isotherm descibes metastable and unstable states  of the system.

The CW equation can also be obtained 
in the grand-canonical ensemble where the magnetic field $h$ is fixed and the total magnetization 
is allowed to fluctuate. One then finds that the Gibbs grand-potential (or "pressure") is given by a variational 
expression $\min_m(\Phi(m) - hm)$. The minimizer satisfies the CW equation with $m$ now interpreted as the average magnetization for a fixed magnetic field $h$. 

One may solve iteratively the CW equation for $m$. When $h>h_{\rm it}>0$ (see fig. \ref{VdW}) one finds a 
unique branch of solutions (global minimizer of $\Phi(m)-hm$). For $h_{\rm it}>h>0$ there appear two new solutions:  
the leftmost solution on fig. \ref{VdW} (local minimizer) and the middle solution on fig. \ref{VdW} (local maximizer). 
Physicaly the global minimizer 
corresponds to a thermal equilibrium state, and the local minimizer (resp. maximizer) correspond to metastable (resp. unstable) states. 
For $h<0$ the situation is symmetrical.

The analogy with coding concepts basically goes as follows.
The magnetic field $h$ is a control parameter analogous to the channel noise $\epsilon$, the Gibbs grand-potential is 
the input-output entropy, the CW equation is the analog of the density evolution fixed point equation, the Van der Waals curve 
the analog of the EBP GEXIT curve,  the iterative threshold $h_{\rm it}$ corresponds to 
$\epsilon_{\rm BP}$ and the static threshold $h_s=0$ corresponds to $\epsilon_{\rm MAP}$.

 \begin{figure}
  \begin{center}
\input{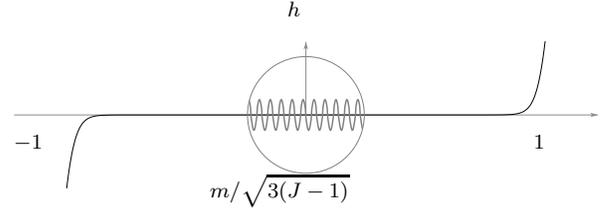}
  \end{center}
  \caption{{\small Van der Waals curve for the coupled CW 
model with $L=25$, $J=1.1$ and $w=1$ (nearest neighbor coupling). 
In the circle the vertical magnification is $10^4$. The iterative threshold nearly coincides with $h_s=0$.}}
  \label{VdWcoupled}
\end{figure}


\subsection{Chain of Coupled Curie-Weiss Systems}

Consider now $2L+1$ copies of the CW system labeled by their "positions" $z$. Thus at position $z$ we have a complete graph with $N$ vertices and spins $s_{iz}$ attached to them.
Each spin at a given position $z\in \{-L+w,...,+L-w\}$ is coupled to $N-1$ spins at the same position with a coupling of strength 
$\frac{J}{2wN}$ and to $2wN$ spins located at positions $z-w,...,-1, +1,...,z+w$ with couplings of strength $\frac{J}{4wN}$. Thus the overall coupling of one spin to the rest of the system is 
$J$ when $N\to +\infty$, just as in the uncoupled system. Close to the boundary, for $z\in \{-L,...,-L+w-1\}$ (resp. $\{L-w+1,...,L\}$) we set couplings towards the left (resp. right) equal to zero and set those to the right (resp. left) equal to $\frac{J}{2wN}$.  The Hamiltonian of the model can be written as
\begin{eqnarray*}
 H_{N,L}(\underline{s}) & = &
-\frac{J}{2wN}\sum_{z=-L}^{L}\sum_{\langle i,j\rangle} s_{iz} s_{jz}
\\ 
& - &\frac{J}{4wN}\sum_{z=-L+w}^{L}\sum_{k=1}^w\sum_{i,j=1}^N s_{i,z-k}s_{jz}\,. \nonumber
\end{eqnarray*}
This is supplemented by a boundary condition: 
for $z\in \{-L,...,-L+w-1\}$ (resp. $\{L-w+1,...,L\}$) the local magnetizations 
are fixed to $\sum_{i} s_{iz} = m_{-}(h)$ (resp. $m_{+}(h)$), where $m_{\pm}(h)$ are the two localy stable solutions 
of the CW equation. In particular, for $\vert h\vert> h_{\rm it}$ they are 
equal (since (\ref{vdw}) has a unique solution) and for $h=h_s=0$ they are opposite. 

\begin{figure}
\input{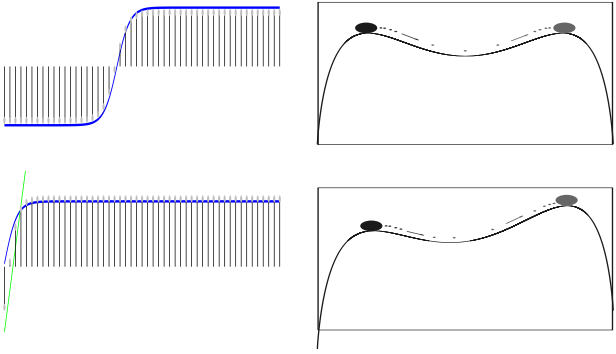}
\caption
{{\small 
{\it Top left:} a kink 
profile for $L=25$,  $J=1.1$, $h=0$; bars 
are the numerical solution $m(z)$ ($z\in\{-25,...,+25\}$) and the continuous blue line is the analytical 
approximation. 
{\it Top right:} a ball rolling 
in the inverted potential $\Phi_h(m)$ (vertical axis), $m$ is the position of the ball (horizontal axis). 
The initial and final velocities are zero and the transition from one maximum 
to the other takes an infinite time.
{\it Bottom left:} profile for $L=25$,  $J=1.1$, $h=0.55$; 
bars are the numerical solution. The continuous blue curve is an analytical approximation 
valid for $z>> -L$, the green segment an analytical approximation 
valid for $z$ near $-L$. {\it Bottom right:} the ball starts from the local maximum with enough initial 
kinetic energy in order to climb on the right maximum of the inverted potential.}
}
\label{inverted_potential}
\end{figure}

This model can be solved exactly either in the canonical or in the grand-canonical ensembles, although the later is technically more convenient. Let $m(z)$ be the average local magnetization
$
m(z) = \bigl\langle \frac{1}{N} \sum_{i=1}^N s_{iz}\bigr\rangle_{N,L}
$
where $\langle - \rangle_{N,L}$ is the average performed with the grand-canonical Gibbs weights $\exp-(H_{N,L}(\underline{s}) + h\sum_{i,z} s_{iz})$. One can derive the following generalization of (\ref{vdw}) for $z=-L+w,...,L-w$
\begin{equation}\label{gvdw}
h= -\frac{J}{4}\frac{D^2}{Dz^2} m(z) - J m(z) + \frac{1}{2}\ln \frac{1+m(z)}{1-m(z)}\, ,
\end{equation}
where $\frac{D^2}{Dz^2}$ is a finite difference operator,
$$
\frac{D^2}{Dz^2} m(z)=\frac{1}{w}\sum_{k=1}^w (m(z-k) - 2 m(z) + m(z+k))\, . 
$$
The solution of (\ref{gvdw}) supplemented with the boundary conditions $m(z) = m_{-}(h)$ for $z\in \{-L,...,-L+w-1\}$ and $m(z) = m_{+}(h)$ for $z\in \{L-w+1,...,L\}$ yields the magnetization profile of the chain. Once this profile is known one can obtain the Van der Waals curve by inverting the relation 
\begin{equation}\label{av}
m = \frac{1}{2L+1}\sum_{z=-L}^{L} m(z)\, .
\end{equation}
As shown in figure \ref{VdWcoupled} this curve has the same characteristics as the GEXIT curve of convolutional LDPC codes. As $L$ 
increases 
the curve becomes almost equal 
to the Maxwell isotherm, already for $w=1$. The wigles correspond to transitions between stable kink states
(see top left in figure \ref{inverted_potential}). As $w$ increases one observes that the amplitude of the wigles diminishes rapidly.

Figure \ref{inverted_potential} depicts the profiles $m(z)$ obtained for 
$L=25$, $w=1$, $J=1.1$. For $h=0$ one finds approximately $50$ stable kink solutions of finite width with a center fixed by 
$m$ as it varies in the region of the wiggles; the figure shows a kink corresponding to $m=0$.
For $h=0.55$ the stable solution of the individual system constitutes 
the major part of the profile.

\subsubsection{A mechanical analogy}
The nature of the solutions $m(z)$ can be understood qualitatively as follows. Defining $\Phi_h(m)= -\Phi(m) + hm$ relation (\ref{gvdw}) can be rewritten as 
$$
\frac{J}{4}\frac{D^2}{Dz^2} m(z) = - \frac{d}{d m(z)}\Phi_h(m(z))\, .
$$
This is a discrete Newton equation for a ball of mass $\frac{J}{4}$ rolling in an {\it inverted} potential $\Phi_h(m)$ (see figure \ref{inverted_potential}). Here $z\in \{-L,...,L\}$ is time and $m(z)$ is the position of the ball. We do not solve the equation of motion for fixed initial position and velocity as is usual in mechanics, but rather for given boundary conditions: we ask that at time $-L$ the ball is at position $m_{-}$ (on the left maximum of the inverted potential) and at time $+L$ it ends up at position $m_{+}$ (on the right maximum). For $h=0$ the initial and final potential energies are the same so that the ball must start and finish with infinitesimal velocity. The ball departs from the left maximum after an infinite amount of time, rolls through the minimum in a finite amount of time, and then climbs to the right maximum in an infinite amount of time.
The instant of departure of the ball is determined by the "average position" $m$, equ. (\ref{av}). On the other hand for $h\neq 0$ the ball starts on the left with finite velocity such that its initial kinetic energy exactly equals the difference in potential energy between the two maxima, because it has to end up on the right with zero velocity. The ball immediately departs from the left, rolls through the minimum in a finite amount of time, and then climbs towards the right maximum in an infinite amount of time. This picture can be confirmed by an explicit analytical solution of the continuum version 
where time $z$ is continuous and the finite difference operator is replaced by a second derivative with respect to time. This calculation yields the following aproximation for the profile when $h=0$,
$$
m(z) \approx \sqrt{3(J-1)} \tanh\bigl\{
\frac{L}{w}\sqrt{\frac{2(J-1)}{J}}
\bigl(\frac{z}{L}+ \frac{m}{\sqrt{3(J-1)}}\bigr)\bigr\}.
$$
This can be shown to be exact in an appropriate scaling limit. For $h\neq 0$ one can also derive separate analytical approximations valid for $z\approx -L$ and $z>>-L$.


\subsection{Chain of Random-Field Curie Weiss Systems}

When an LDPC code is viewed as a spin system the channel outputs play the role of a random magnetic field that is added to the Hamiltonian. This is our motivation to consider the RFCW model
and check that the general picture of the previous paragraph still holds. The RFCW model is defined by adding
a contribution $-\sum_{iz} H_{iz} s_{iz}$ to the Hamiltonian
where $H_{iz}$ are i.i.d random variables 
with a well behaved symmetric density ${\bf E}[H_{iz}]=0$. 
A generalized form of coupled Curie-Weiss 
equations determines a profile for the expected 
value of the magnetization and a Van der Waals 
curve can again be defined from 
(\ref{av}). 
Numerical solutions display the same features 
as in the deterministic case.
We do not give more details here due to lack of space.


\section{Coupled Constraint Satisfaction Models}

We now turn to a much more challenging class of models, 
namely constraint satisfaction models defined on random sparse graphs, and  
focus on two paradigms, $K$-satisfiability and $Q$-coloring. 
Both models display a static sat-unsat phase transition 
as a function of a control parameter\footnote{As we will see, for $K$-SAT $p=\alpha$ the clause density and 
for $Q$-COL $p=c$ where $c/N$ is the probability that an edge is present.}
$p$: for $p<p_s$ one finds exponentialy many solutions satisfying {\it all} constraints (sat phase), while for $p>p_s$ no such solutions exist (unsat phase).
This is the prediction of the zero temperature cavity method 
applied to locally tree like graphs \cite{Mezard-Parisi}. Within 
this formalism 
the sat and unsat phases are the globaly stable solutions of a 
set of fixed point equations called survey propagation (SP) equations. 
SP equations are 
a set of message passing equations - which may also be viewed 
as BP equations 
associated to a ``derived graphical model'' - where the 
messages are probabilities attached to the edges of the graph. 
The sat phase is associated to a {\it trivial}
fixed point solution\footnote{Strictly speaking one looks at 
the fixed point equation satisfied by the probability densities of 
the messages.} that exists for all $p$. 
To find the other solutions one solves the SP equations 
iteratively and it turns out that when $p$ crosses
$p_{\rm SP}$ ($p_{\rm SP}\leq p_s$), called the SP threshold, non 
trivial fixed points appear. In $[p_{SP},p_s]$ the 
geometry of the 
space of solutions is non trivial and may 
display various other thresholds, a question that we do not adress here. 

We will show, both numericaly 
and analyticaly (in large $K$ and $Q$ regimes) that coupling 
again leads to the boosting 
of the message passing threshold $p_{\rm SP}^{\rm coupled}\to p_s$. 
As expected one also finds
$p_{s}^{\rm coupled}\to p_s$.
It is 
natural to conjecture that other thresholds of 
the coupled model that may
exist in $[p_{\rm SP}^{\rm coupled},p_{s}^{\rm coupled}]$ 
also tend to $p_s$.

\subsection{$K$-Satisfiability}

\subsubsection{General formalism}
Consider a general but fixed bipartite graph $G$ with variable nodes 
 $\{i,j,...\}$, clause nodes $\{a, b, ...\}$, and edges 
connecting only variable to clause nodes. Edges come in two 
versions, dashed or full. Dashed edges are marked with 
$J_{ia}=1$ and full ones with $J_{ia}=-1$. The set of neighboring nodes of $a$ (resp. $i$) is called 
$V_a$ (resp. $V_i$). Boolean variables $x_i\in\{0,1\}$ are assigned 
to variable nodes and a clause $a$ is 
said to be satisfied iff $\vee_{i\in V_a}
x_i^{J_{ia}} =1$, where $x_i^{+1}=x_i$ and $x_i^{-1}=\bar{x}_i$. 
A satisfying assignement is one that
satisfies all clauses simultaneously $\wedge_a(\vee_{i\in V_a}
x_i^{J_{ia}}) =1$. Note that this model 
can easily be cast in Hamiltonian form in terms of Ising spins.

We briefly explain the content of SP message passing equations. 
Here we adopt the formalism of  
\cite{Mertens-Mezard-Zecchina}.
If $a$ is unsatisfied by all variables in  $V_{a}\setminus i$ then it sends a "warning" to $i$ with probability $\eta_{a \to i}$. This is 
computed from other probabilities sent to $a$ by  nodes $j\in V_{a}\setminus i$. Consider the warnings received by $j$ from clauses $b\in V_j\setminus a$. These come in two categories: those that are "impeding $a$" and those that are  "supporting $a$". A warning received by $j$ impedes $a$ when the edges 
$(b\to j\,;\,j\to a)$ are (dashed; full) or (full; dashed) 
because $j$ will obviously 
not be able to satisfy both $a$ and $b$. 
On the other hand if the edges $(b\to j\,;\,j\to a)$ are (dashed; dashed)
 or (full; full) the warning received by 
$j$ supports $a$. 
Let $\pi_{j\to a}^{+}$ 
(resp. $\pi_{j\to a}^{-}$) be the 
probabilities that $j$ receives  
no supporting (resp. no impeding) warning. We 
have:
\begin{eqnarray*}
\pi_{j\to a}^{\pm} & = & \prod_{b\in V_{ia}^{\pm}} (1-\eta_{b\to j}), \qquad j\in V_a\setminus i \\
\eta_{a\to i} & = & \prod_{j\in V_a\setminus i}
\frac{\pi_{j\to a}^{+} (1-\pi_{j\to a}^{-})}{\pi_{j\to a}^{+}
+ \pi_{j\to a}^{-} - \pi_{j\to a}^{+}\pi_{j\to a}^{-}}
\end{eqnarray*}
The two disjoint sets $V_{ia}^{+}$ (resp.  $V_{ia}^{-}$) are such that edges $(b\to j\,;\,j\to a)$ are of the same (resp. different) type. 


\subsubsection{Coupled $K$-SAT}

We apply this formalism to graphs that are instances 
of a coupled ensemble. 
At each position $z=-L-w+1,...+L+w-1$ we lay down sets of
clauses $a_z$ and variable nodes $i_z$ whose cardinalities have 
ratio $\alpha$. 
Each clause $a_z$ has $K$ emanating edges. Each edge 
connects to a node at position $z+k$, call it $i_{z+k}$, where $k$ is picked 
uniformly at random in $0,...,w-1$ and then
$i_{z+k}$ is picked uniformly at random from nodes at the position $z+k$. An edge is then turned into dashed or full with probability $\frac{1}{2}$.
Note that the degree of the clauses is equal to $K$ and the one of the variable nodes is ${\rm Poisson}(\alpha K)$. Note also that for 
$w=1$ we recover the usual uncoupled $K$-sat graphs.
 
The distributional 
equations associated with 
the message passing system are solved 
iteratively by a sampled density evolution or  
population dynamics method. It is convenient to work with the entropic
variables $\phi=-\ln(1-\eta)$ and $x^\pm = -\ln \pi^\pm$. There always exist a trivial
fixed point ${\bf E}[\phi]$ = 0 corresponding to 
a vanishing ``warning entropy''.
Below a treshold $\alpha_{\rm SP}^{\rm coupled}$ this is the 
unique fixed point, and non trivial solutions 
for the warning entropy appear above this threshold. 
Let us list some of our numerical observations.
For $K=3$ we 
have $\alpha_{\rm SP}=3.93$, $\alpha_s=4.266$ and 
$\alpha_{\rm SP}^{\rm coupled}(L=30, w=3)=4.270$, 
$\alpha_{\rm SP}^{\rm coupled}(L=40, w=3)=4.268$. 
For $K=4$ we have $\alpha_{\rm SP}=8.3$, $\alpha_s=9.931$ and 
$\alpha_{\rm SP}^{\rm coupled}(L=30, w=3)=9.935$,
$\alpha_{\rm SP}^{\rm coupled}(L=50, w=3)=9.932$. The message passing threshold 
of the chain comes very close to $\alpha_s$ already for small values of $w$.

\subsubsection{Large $K$-limit}

For the uncoupled system in this limit $\alpha_s= 2^K\ln 2 - \frac{1+\ln 2}{2}$ and 
$\alpha_{\rm SP}= 
\frac{{2^K}}{K}(\ln K +\ln\ln K +1-\ln 2 + O(\frac{\ln\ln K}{\ln K}))$ 
\cite{Mertens-Mezard-Zecchina}.
Thus we set $\alpha = 2^{K}\widehat\alpha$, and take $K$ large while $w$ and $L$ are kept fixed. 
In terms of the entropic variables the message 
passing equations become sums with $O(K)$ terms and one can reasonably assume that their distibution is picked
on their average values. Using this approximation we find
for $z\in -L,...,L$
$$
\varphi(z) = \widehat\alpha K 
\biggl\{\frac{1}{w}\sum_{k=0}^{w-1}
\frac{e^{\frac{1}{w}\sum_{j=0}^{w-1}\varphi(z-j+k)}-1}
{e^{\frac{1}{w}\sum_{j=0}^{w-1}\varphi(z-j+k)}-\frac{1}{2}}\biggr\}^{K-1}
$$
where $\varphi(z) = 2^{K-1} \widehat\alpha K{\bf E}[\phi_z]$ is 
the scaled warning entropy
emanating from nodes at position $z$. 
For $w=1$ we get the 
equation satisfied by the individual uncoupled system, namely
$
\varphi = \widehat\alpha K 
\{\frac{e^{\varphi}-1}{e^{\varphi}-\frac{1}{2}}\}^{K-1}
$.
This equation is the analog of the 
Van der Waals curve (\ref{vdw}) (or EBP GEXIT function), and serves to fix the boundary 
conditions in the set of coupled equations. 
Given $\alpha$, for $z\notin\{-L,...,L\}$ we fix  
$\varphi(z)$ to the trivial fixed point solution on the left
and to the other non-trivial globaly stable fixed point on the right.
\begin{figure}
\centering
\input{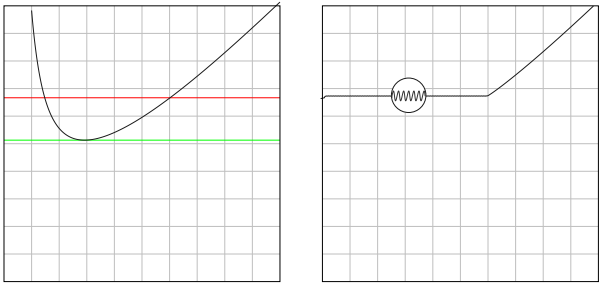}
\caption{{\small 
VdW curves for $5$-SAT.
Vertical axis is $\widehat\alpha$ and 
horizontal is the average warning entropy.
{\it Left:} uncoupled $5$-SAT;
the upper red line is 
at $\widehat\alpha_s=\ln 2 - \frac{1+\ln 2}{64}\approx 0.6666$ 
and the lower green 
line is at $\alpha_{\rm SP} \approx 0.5129$. 
{\it Right:} $(L=50, w=3)$. In the circle the vertical magnification is $10^2$. There are approximately $100$ wigles corresponding to 
transitions between kink states for the profile of the warning entropy. As $w$ increases their amplitude rapidly diminishes.}
}
\label{ksatVdW}
\end{figure}
The Van der Waals curve for the coupled system is given here by the plot of 
$\widehat\alpha$ as a function of $\overline{\varphi}$ where
$\overline\varphi=\frac{1}{2L+1}\sum_{z=-L}^L\varphi_z$ (see figure \ref{ksatVdW} for numerical solutions).

\subsection{$Q$-Coloring}

In the $Q$-coloring problem one assigns 
colors among $\{1,...,Q\}$ to all the vertices of a graph. 
Assignements such that any pair of adjacent nodes have different colors are 
called satisfying assignements. 
The problem is equivalent to the Potts model Hamiltonian at 
zero temperature and it has been analyzed by the 
cavity method for instances of Erd\"os-R\'enyi  
random graphs \cite{Mulet-Pagnani-Weigt-Zecchina}. 
This leads to a set of SP equations.  

We analyze the SP equations for an ensemble of Erdoes-R\'enyi graphs 
coupled into a one dimensional chain. 
At each position $z\in\{-L, ...,L\}$ 
place a collection of vertices $i_z\in \{1,...,N\}$. 
Now for each pair of positions such that
$\vert z_1 - z_2 \vert\leq 2w$ (we allow $z_1=z_2$) 
consider all pairs of distinct vertices $i_{z_1}j_{z_2}$ and 
connect them with an edge, 
with probability $\frac{c}{2w+1}\frac{1}{N}$. Given a 
position $z$ in the bulk
and a vertex $i_z$ there are $d_{i_z} \sim \textrm{Poisson}(c)$ 
outgoing edges all connected to vertices that are at positions in 
the window $\{z-w,...,z+w\}$.
The case $w=0$ corersponds to the individual uncoupled system. 

The system displays phase transitions as a function of the graph 
density $c$ with the same numerical features as in $K$-SAT. 
An analytical analysis for the limit of large $Q$ 
confirms the numerical findings.


\section{Discussion}

In the ususal paradigms of statistical 
mechanics, LDPC and other coding constructions, or random constraint satisfaction problems, one looks at the macroscopic 
behavior of {\it large assemblies of interacting microscopic sytems}.
Here we look at the 
behavior of {\it large assemblies of mean field macroscopic systems 
coupled into a one-dimensional chain}. The equilibrium phases of the individual 
system are 
forced to coexist by the boundary
conditions. This generates the kink solutions that appear as 
stable solutions 
of the fixed point equations describing the chain. 
It is the stability of these solutions that
implies the saturation of the dynamic threshold of the coupled ensemble to 
the static one of the single system. Clearly this mecanism can work out 
only if the single system has no spatial 
or finite-dimensional geometrical structure (i.e. is mean field).
Indeed  
any system defined on ${\bf Z}^d$ can be thought as a coupled chain of
systems on ${\bf Z}^{d-1}$; but we know that the 
phenomenon of threshold saturation does not occur for such systems (e.g finite dimensional Ising models).

Our systems are a hybrid between mean field models and Kac models.
In Kac models degrees of freedom (e.g spins, atomic positions) live in a finite dimensional space and 
interact via a finite range pair potential, whose length 
scale is sent to infinity while its intensity is sent to zero in a proper way. 
In the limit, the Kac model Hamiltonian looses all trace of spatial 
structure and becomes mean 
field (e.g it may ressemble the CW Hamiltonian). 
But because the free energy of a system with local 
potential has to be convex, the free energy of the 
Kac model is still convex in the limit. This means that it will be 
given exactly by the Maxwell construction (or convex hull)
applied to the one of the mean field limiting Hamiltonian. This was 
proven in the classic work \cite{Lebowitz-Penrose}.
Now, the 
convolutional LDPC ensembles, as well as the other models discussed in this note, have Hamiltonians that are
already mean field in the direction "perpendicular" to the chain, and have a Kac interaction of width $O(w)$ 
along the direction ``parallel'' to the chain. 
Thus one would expect that as $w$ grows the GEXIT/Van der Waals curves
of the coupled chain comes close to the Maxwell construction for 
the limiting model. And since the limiting model is ``just a bigger'' 
uncoupled model, the GEXIT/Van der Waals curve of the couped system should 
converge to that of the individual mean field system. 
These arguments show that threshold saturation should be expected.

Clearly there are many open questions that are worth investigating. Here let us just
mention possible connections with coarse grained theories of interfaces,
coupled map systems, discrete soliton equations, and perhaps most importantly algorithmic 
implications of the phenomenon of treshold saturation.



\section*{Acknowledgments}

{\small We acknowledge instructive discussions with S. Kudekar, and A. Montanari.
The work of S. H. Hassani was supported by grant no 200021-121903 of the Swiss National Science Foundation.}





\end{document}